# Superhydrophobic/superoleophilic magnetic elastomers by laser ablation

Athanasios Milionis [a,*], Despina Fragouli [a], Fernando Brandi [a], Ioannis Liakos [a], Suset Barroso [a], Roberta Ruffilli [b], Athanassia Athanassiou [a],*

a Smart Materials-Nanophysics, Istituto Italiano di Tecnologia (IIT), Via Morego 30, 16163 Genova, Italy

b Nanochemistry, Istituto Italiano di Tecnologia (IIT), Via Morego 30, 16163 Genova, Italy

E-mail addresses: am2vy@virginia.edu (A. Milionis), athanassia.athanassiou@iit.it (A. Athanassiou).

**Abstract**

We report the development of magnetic nanocomposite sheets with superhydrophobic and supe-oleophilic surfaces generated by laser ablation. Polydimethylsiloxane elastomer free-standing films, loaded homogeneously with 2% wt. carbon coated iron nanoparticles, were ablated by UV (248 nm), nanosecond laser pulses. The laser irradiation induces chemical and structural changes (both in micro-and nano-scale) to the surfaces of the nanocomposites rendering them superhydrophobic. The use of nanoparticles increases the UV light absorption efficiency of the nanocomposite samples, and thus facilitates the ablation process, since the number of pulses and the laser fluence required are greatly reduced compared to the bare polymer. Additionally the magnetic nanoparticles enhance significantly the super-hydrophobic and oleophilic properties of the PDMS sheets, and provide to PDMS magnetic properties making possible its actuation by a weak external magnetic field. These nanocomposite elastomers can be considered for applications requiring magnetic MEMS for the controlled separation of liquids.

## 1. Introduction

Superhydrophobic surfaces with apparent water contact angle (APCA) greater than 150° have gathered significant attention during the past years, due to their suitability for various applications such as self-cleaning surfaces [1], sensors [2], manipulation of water droplets [3], anti-icing [4] and anti-fouling surfaces [5,6], etc. More recently, there is an enhanced interest for superhydrophobic mate-rials that exhibit simultaneously superoleophilic properties. Such materials in the form of sponges [7,8] or membranes [9] are very attractive for separating large volumes of oil–water mixtures. These approaches are based on application of diverse particles or chemical substances by electrostatic deposition [7,10], absorption or colloidal dispersions by capillary flow [7], dip-coating [8,11] or phase-inversion [9,12], techniques that work very well when a material with homogeneous surface roughness and chemistry is desired. In addition, if these materials exhibit magnetic properties on top of their unique surface characteristics, they can be actuated or manipulated by external magnetic fields and eventually can be attractive for sophisticated applications requiring smart material behavior [7,13]. However, for the fabrication of functional micro-devices (e.g. lab-on-a-chip, microfluidics) [14,15] where the surface characteristics have to be spatially tuned in order to interact differently with various types of liquids, these techniques cannot be a potential solution and another fabrication approach has to be considered.

So far, controlling spatially the surface chemistry and/or rough-ness has been achieved by spray deposition of colloidal suspensions through shadow masks [3], pulsed [16] and continuous laser irradiation [17], etc. In other cases electrical external stimuli are used to induce in situ wetting gradients [18]. Although these studies can successfully create wetting patterns, they usually study the interaction of one type of liquid, while a combined superhydropho-bic/superoleophilic localized effect is not reported.

Among the materials that have been used for the realization of superhydrophobic surfaces, polydimethylsiloxane (PDMS) presents diverse advantages. It is an elastomeric organosilicon compound, chemically inert, non-toxic and non-flammable and has been widely used in soft-lithography due to its unique rheological properties for the fabrication of microfluidic devices, sensors, flexible electronics, biotechnological devices [19,20], etc. Since PDMS is intrinsically a hydrophobic and oleophilic polymer, increasing its surface roughness is expected to enhance the wetting properties of its surface resulting in the realization of superhydrophobic and superoleophilic materials.

For the surface modification of PDMS (in terms of topography and surface chemistry) different techniques have been followed so far including soft molding [21–23], acidic treatments [24], plasma etching [25], laser micromachining [26–30] or combinations of these techniques [31]. Alternatively, surface modification of PDMS can be achieved by mixing it with nanomaterials. Basu et al. and Zhao et al. have enriched the PDMS matrix with silica nanofillers in order to form a nanocomposite material that after spraying resulted in a surface with enhanced roughness and superhydrophobic properties [32,33]. In addition, Nagappan et al. and Wang et al. have examined dip-coating with nanocomposite solutions as a potential strategy to obtain superhydrophobic materials [34,35]. Finally, Tropmann et al. have followed a plasma etching approach to render PDMS/poly(tetrafluoroethylene) nanocomposites superhydrophobic [36]. Among them, only Tropmann et al. [36] locally altered the surface properties of PDMS nanocomposites but the procedure was time consuming and required multiple fabrication steps. Com-pared to the aforementioned techniques, laser treatment methods offer significant advantages when it is required repeatable, precise and highly controlled surface modification [37]. Towards this direction Villafiorita-Monteleone et al. have used cycles of pulsed UV laser irradiation and vacuum storage on PDMS samples filled with organic-capped nanorods of $TiO_2$ to induce reversible sur-face transitions to the nanocomposites between hydrophobic and hydrophilic states, for the realization of microfluidic devices with controlled liquid flow [38]. In that case the laser fluence was kept low enough in order to avoid any topography changes on the sur-face of the nanocomposites and only the intrinsic property of $TiO_2$ that becomes hydrophilic upon UV irradiation was exploited. By properly adjusting parameters such as laser fluence, but also pulse duration, repetition rate and wavelength of irradiation it is pos-sible to induce different local effects on the surface topography [39]. Combining such a laser fabrication approach with the addition of nanofillers that exhibit desired magnetic, electric, mechanical or other properties, we can envision the fabrication of materials with controlled surface chemistry characteristics combined with improved bulk properties.

Up to now, fabrication of superhydrophobic PDMS surfaces by direct laser irradiation has been achieved by using Nd:YAG lasers operating in the second harmonic (532 nm) [26], femtosecond lasers operating in the infrared region [27,28], $CO_2$ lasers [40] or UV excimer lasers [41–43]. All of the aforementioned studies investigate the surface roughening of pure PDMS and none of them reports surface modification of novel PDMS nanocomposites with enhanced surface and bulk properties. The nanocomposites fabricated in this work, require lower laser fluence compared to pure PDMS, in order to change their surface topography by ablation, and eventually to obtain the desired superhydrophobic/superoleophilic properties. Additional they exhibit magnetic actuation properties.

In particular, we report the fabrication of superhydropho-bic/superoleophilic magnetic nanocomposite thin sheets using ablation induced by a UV excimer laser. We investigate how the incorporation of magnetic nanoparticles (NPs) in PDMS can improve the fabrication process and also provide additional unique properties to the material. The magnetic NPs utilized as fillers in the PDMS matrix exhibit ferromagnetic properties [44]. Therefore, the presence of these NPs introduces magnetic properties to the films. The resulting nanocomposites after the laser treatment exhibit significantly improved properties compared to the ablated bare polymer in terms of hydrophobicity, and oleophilicity, while the ablation threshold of the former is significantly lower. In summary, this technique offers localized surface modification with superhydrophobic and superoleophilic properties while the uniqueness of the hydrophobically capped magnetic nanoparticles, like the ones used here, lies in the fact that they can: (1) reduce the fabrication cost (by improving the laser light absorption due to their wide absorption spectra and therefore reduce the number of pulses required for superhydrophobicity/superoleophobicity), (2) induce magnetic actuation due to their ferromagnetic properties,(3) disperse well in non-polar solvents for the sample preparation, and (4) provide nanoscale surface roughness that enhances the hydrophobicity and oleophilicity. These functional films apart from their enhanced magnetic properties, are also durable after repeated cycles of mechanical abrasion. Such multifunctional systems can find potential applications in MEMS, sensors and sophisticated technological devices.

**2. Materials and methods**

2.1. Materials

Spherical ferromagnetic iron NPs with hydrophobic carbon shell of an average particle size 30–60 nm were purchased from Plasmachem, Germany. PDMS (Sylgard 184 Silicone Elastomer) purchased from Dow Corning Corporation was supplied in two compounds: a pre-polymer and a crosslinker. All solvents used were purchased from Sigma Aldrich.

2.2. Samples preparation

The bare PDMS surfaces were prepared by mixing the prepolymer with the cross-linker in a 10:1 weight ratio and subsequently cured at 80 ◦C for 1 h in a conventional oven. To prepare magnetic field-responsive PDMS composites, a quantity of iron NPs was dispersed homogenously in chloroform after sonication for 20 min.

The prepolymer was mixed with an aliquot of the NPs solution in a weight ratio 98:2 (prepolymer:NPs) and was left under stirring for a few minutes until a homogeneous dispersion was obtained. The nanocomposite to chloroform weight ratio was kept constant 1:1 in all the experiments. Finally, the crosslinker was mixed with the nanocomposite solution in a 1:5 mixing ratio (e.g. 1 g of crosslinker for every 5 g of nanocomposite prepolymer solution). In order to facilitate the peeling off of the final films, the final solution was casted on PTFE substrates and left under the fume hood until complete solvent evaporation. Subsequently, the samples were cured at 70 °C for 3 h. With these conditions, it was possible to fabricate films with approximately 400 m thickness measured by a micrometer.

2.3. Laser ablation

The peeled-off sheets were ablated with a pulsed KrF excimer laser (Coherent-CompexPro 110) operating at 248 nm wavelength with pulse duration 20 ns and repetition rate 20 Hz. The samples were irradiated at constant laser fluence (F) 0.8 J/cm$^2$ (unless otherwise written) and different number of pulses. A beam expander and a square diaphragm were used to select the most homogeneous part of the laser beam that was focused onto the surface of the sample with a projection lens. A x–y motorized translational stage con-trolled by computer and positioned perpendicularly to the direction of the laser beam was used to move the samples and perform a step-and-repeat laser scanning. In this way, large surface areas of several cm2 could be ablated. Finally, nitrogen flow was used during the ablation to avoid the re-deposition of the removed material. After the laser treatment, the samples were left in ambient conditions for 48 h before any characterization. The laser-treated films could be rendered porous through the fabrication of laser-drilled apertures of different diameters by using a laser fluence of 7 J/cm$^2$ and 1700 pulses. Different size of apertures could be obtained by changing the size of the irradiated area while maintaining the same expo-sure conditions (fluence and number of pulses). The laser exposure occurred on both sides of the samples (on top and below).

2.4. Characterization of the samples

Measurements of the APCA of water and oil, as defined by Marmur [45], were carried out using an OCAH 200 video based optical contact angle measuring instrument, (DataPhysics, Germany). Mul-tiple measurements were taken by gently depositing 5 l drops of deionized water and mineral oil on several samples prepared with the same procedure. The standard error for the wetting measurements was ±4°. The morphology of the laser patterned surfaces was characterized by a variable pressure Jeol JSM-6490LA scanning electron microscope (SEM). For extracting surface rough-ness parameters a Park Systems XE-100 atomic force microscope (AFM) was used in non-contact mode with a silicon cantilever. X-ray photoelectron spectroscopy (XPS) measurements were conducted using a Specs Lab2 electron spectrometer equipped with a monochromatic Mg K X-ray source radiation at 1.253 eV and with a Phoibos hemispherical analyzer Has 3500. Finally, the photographs of the samples were acquired with a Canon EOS 5D Mark II camera equipped with a Canon EF 100 mm f/2.8 l IS USM Macro objective lens.

**3. Results and discussion**

Fig. 1 depicts optical images of the bare and laser treated PDMS (Fig. 1a) and nanocomposite (Fig. 1b) sheets. The laser treated areas were obtained after irradiation with 30 pulses, and fluence of 0.8 J/cm$^2$ on adjacent rectangular areas of 4.41 mm$^2$. As shown, both the bare PDMS and the nanocomposite samples change color after laser ablation, with the first turning into a yellowish color and the second into light gray.

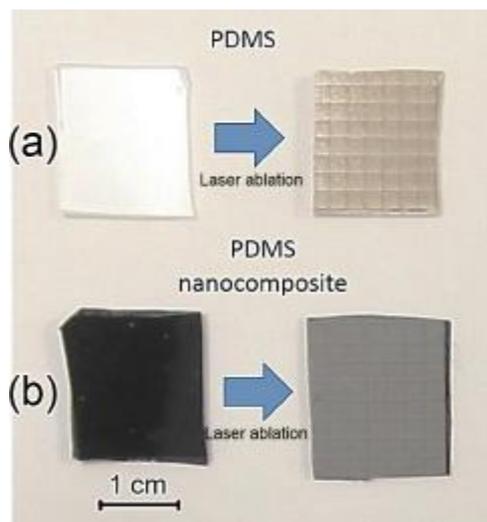

Fig. 1. (a) Bare PDMS before (left) and after (right) laser-ablation. (b) NPs-PDMS film before (left) and after (right) laser-ablation. The laser-treated samples were exposed to 30 pulses of 0.8 J/cm$^2$ fluence.

For identical irradiation conditions, the two materials exhibit significantly different wetting properties. As shown at Fig. 2a, the bare PDMS and the nanocomposite surfaces have quite similar APCA (119 ± 4° and 120 ± 4°, respectively) before irradiation. However, after laser irradiation their wetting behavior is clearly differentiated. In particular, the ablated nanocomposite surfaces reach the superhydrophobic behavior after irradiation with 15 laser pulses at fluence 0.8 J/cm$^2$, with APCAs greater than 150° and roll off angles less than 5°, whereas the APCA of bare PDMS is slightly increased under the same conditions, staying far from the superhydrophobic region even after 30 pulses. Fig. 2b shows the rolling state of a water drop on the samples irradiated with 15 pulses. The bare PDMS sheets need to be irradiated with much higher fluence (2 J/cm$^2$) in order to become superhydrophobic after 20 incident pulses. This indicates that the presence of NPs greatly facilitates the achievement of extreme wetting behavior in the ablated elastomer sheets (Video 1 Supporting Info). The used concentration of NPs (2% wt.) that was kept stable throughout all the experiments is sufficient to provide magnetic actuation properties to the samples. In Fig. 3 it is demonstrated the magnetic manipulaion of the samples by using a 0.5 T magnet. As the magnet is moved towards the sample, the magnetic response of the nanocomposite sheet which is attracted from the magnet can be clearly seen. Video 2 of the Supporting Info shows a video with magnetic actuation of the samples.

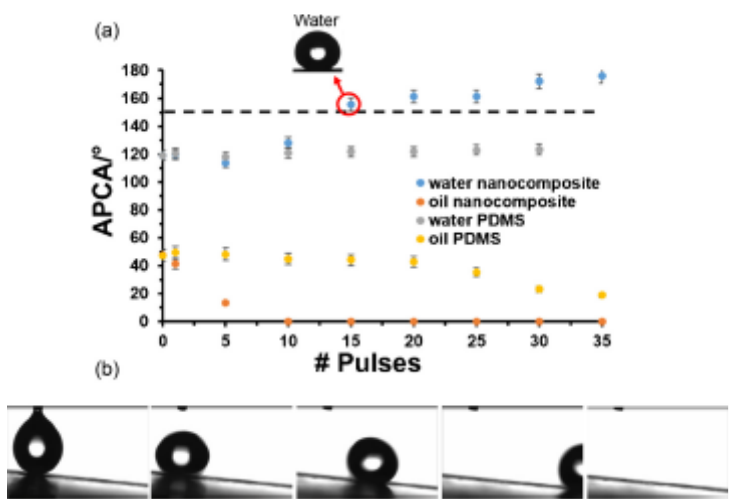

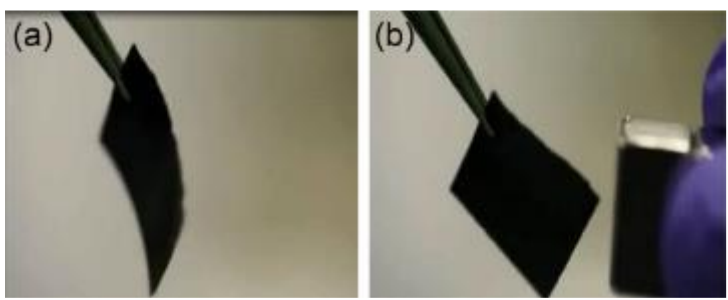

Fig. 2. (a) APCA of water and oil vs. delivered laser pulses on the NPs-PDMS and PDMS sheets surface. The laser fluence used is 0.8 J/cm$^2$. Inset: Water APCA for the NPs-PDMS after 15 UV pulses. (b) Rolling state for water drop on the laser-ablated NPs-PDMS sheets with 15 pulses at tilt angle of 5°.

Fig. 3. (a) The nanocomposite elastomer is kept hold by a tweezer. (b) The magnet approaches and attracts the elastomer.

Additionally, the nanocomposite samples were found to exhibit complete oil wetting (oil APCA: 0°) after just 10 pulses at 0.8 J/cm$^2$,(superoleophilic), whereas the respective value for the non-irradiated samples was oil APCA: 48 ± 4° (Fig. 2a and Video 3 Supporting Info). The pure PDMS samples show an oil APCA of 47 ± 4° before laser irradiation, whereas after laser irradiation with 35 pulses of 0.8 J/cm$^2$ laser fluence, the oil APCA decreases down to 19 ± 4°. It is worth noting that the performance of the fabricated samples in terms of superhydrophobicity and superoleophilicity was not degraded in time, since the samples were tested several times during a period of several weeks.

Surface topography and chemistry are the two main com-ponents that affect the surface wettability, and for this reason extensive analysis of both follows. In order to evaluate and explain how the surface morphology influences the APCA, the SEM analysis of nanocomposite films irradiated with different number of pulses was conducted. As shown in Fig. 4, the exposure to increasing number of laser pulses enhances significantly the surface roughness of the nanocomposites. In fact, a multi-scale rough surface with nano-, submicro- and micro-features is gradually formed, leading to the representative surface morphology of Fig. 4e, for which the superhydrophobicity is achieved. The pulses delivered on the nanocomposites cause the gradual formation of micro-holes that become denser with increasing number of pulses (Fig. 4a–c). This micro-hole formation on laser irradiated PDMS has been reported in the literature by Graubner et al. as local chemical transformations that take place in the initial incubation phase and they were attributed to impurities remaining from the synthesis which act as "seeds" for these transformations [42]. After approximately 15 pulses (0.8 J/cm2) the micro-holes start to form a continuous network that renders the surface superhydrophobic. This value is significantly lower compared to the study of van Pelt et al. on PDMS substrates where they used at least 3 times higher total energy delivered to the sample in order to render it superhydrophobic (70 pulses with average laser fluence 547 mJ/cm$^2$) [41]. Moreover, the laser fluence used here to render superhydrophobic the nanocomposite sheets is lower than the ablation threshold of PDMS that was reported as 0.94 J/cm2 (after 400 pulses) by Graubner et al. [42]. In the present case, after only 30 pulses (0.8 mJ/cm$^2$) a well-organized multi-scale roughness surface is formed (Fig. 4d,e), responsible for the resulting superhydrophobic and superoleophilic properties.

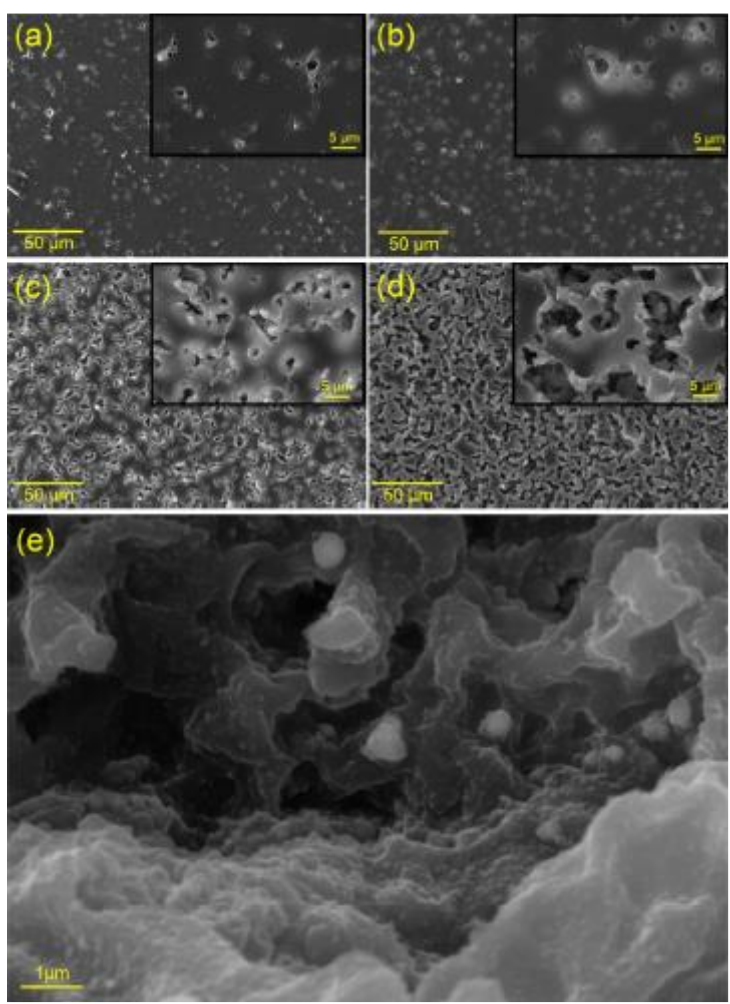

Fig. 4. SEM images of NPs-PDMS surfaces treated with (a) 1, (b) 3, (c) 9, and (d) 30 laser pulses. Higher magnifications of the obtained structures are shown in the corresponding insets. (e) High magnification image showing details of the surface of the sample irradiated with 30 pulses. In all cases F = 0.8 J/cm$^2$.

The contribution of the NPs in the absorption of the laser irradiation is significant since they absorb in all the UV region as it is clearly shown from their absorption spectra obtained by a UV spectrophotometer (Figure 1 in Supporting Information). Moreover it is known that nanoparticles can generate heat upon laser irradiation. This may lead to physicochemical modifications in their surrounding areas that can alter the wetting properties [46,47]. On the other hand PDMS absorbs only in the deep UV region.

An atomic force microscope was used to extract surface rough-ness parameters and evaluate the etching depth after 30 pulsed. The etching depth of the irradiated nanocomposite material was approximately 3 μm. Surface parameters that were estimated include the arithmetic mean, the root mean squared, the skewness and the kurtosis. In particular, the arithmetic mean was measured 0.165 μm and the root mean squared 0.915 μm while the skewness was measured 1.174 and the kurtosis 4.841 that correspond to a relatively spiky surface with high peaks and deep valleys (leptokurtic). Fig. 5 shows the actual profile of the surface as was recorded by AFM. The scanning was performed on the border between ablated and non-ablated region in order to show clearly the difference of the surface morphology.

In order to evaluate the surface chemistry modifications that take place upon laser treatment of the surfaces, and their possible contribution to the induced superhydrophobicity, the samples were investigated with XPS (Table 1). From the wide scan, it is observed that there is a decrease in the concentration of the C atoms when the surfaces are exposed to the laser pulses, accompanied by an increase in the elemental concentration of Si atoms, explaining the increase in the APCA, since the organosilicon compounds are known for their hydrophobicity [48]. In particular, in the bare PDMS, irradiated with 30 pulses at 0.8 J/cm2, the initial surface concentration of the C atoms is 46.85% and after the laser-treatment falls to 43.07% (decrement 3.78%, see Table 1), while the Si slightly increases by 2.22%. These small variations before and after the laser irradiation may be responsible for the small increase of the APCA, shown in Fig. 1. However, in the case of the nanocomposite sheets, irradiated with the same conditions, the C reduction is higher (9.43%), while the presence of the Si atoms is increased by 3.37%, indicating a more drastic modification on the surface of the ablated material. The decrease of the exposed carbon observed, can be attributed to the removal of organic compounds due to the laser ablation, but also due to a possible chemical modification of the surface. It is worth noting that the XPS analysis of the NPs pow-der, shows that the surface of the NPs is mainly covered by a layer of carbon (93%) while the iron element is slightly evident (0.65%)(Table 1). For this reason, when the NPs were incorporated into the PDMS, either before or after ablation no Fe2p was detected on the outermost surface (Table 1) since most probably the signal is hindered by the capping molecules of the NPs and the polymer matrix; taking into account that the XPS analysis depth is equal or less than 10 nm.

Additionally, as confirmed by the C1s high resolution spectra, the PDMS matrix was characterized by C–Si & C–H bonding typical of the polymer, in accordance with other studies (Sup-porting Information Fig. 2a) [49]. C1s high resolution scans were acquired also for the other surfaces, i.e. after laser ablation (Sup-porting Information Fig. 2b) or after NPs incorporation (Supporting Information Fig. 3a) and showed no change in the C1s bond chemistry. On the contrary, when the polymer nanocomposite was laser micromachined, an additional component in the C1s region appeared, shifted by approximately 1.2 eV higher binding energy than the CSi, CC, and CH bonds (Supporting Information Fig. 3b). This new component could be attributed to oxidized carbon (Sup-porting Info Fig. 4) present in the hydrophobic capping molecules of the Fe NPs. To prove this, the XPS C1s region of bare Fe

NPs was acquired and revealed similar carbon chemistry as the laser micromachined nanocomposite (Supporting Information Fig. 4). This experiment demonstrates therefore, that the hydrophobic cap-ping molecules get exposed to the outermost surface (10 nm) of the laser ablated polymer nanocomposites.

Thus, we may conclude that the superhydrophobic/superoleophilic properties of the laser ablated nanocomposites arise from the combination of two different factors. First, from the increase of the surface roughness due to the laser ablation and consequently the enhanced contribution of the hydrophobically capped magnetic NPs (Fig. 4e). Second, from the increased presence of the hydrophobic–Si-O-Si- skeletal structures on the nanocomposites' surfaces after laser ablation.

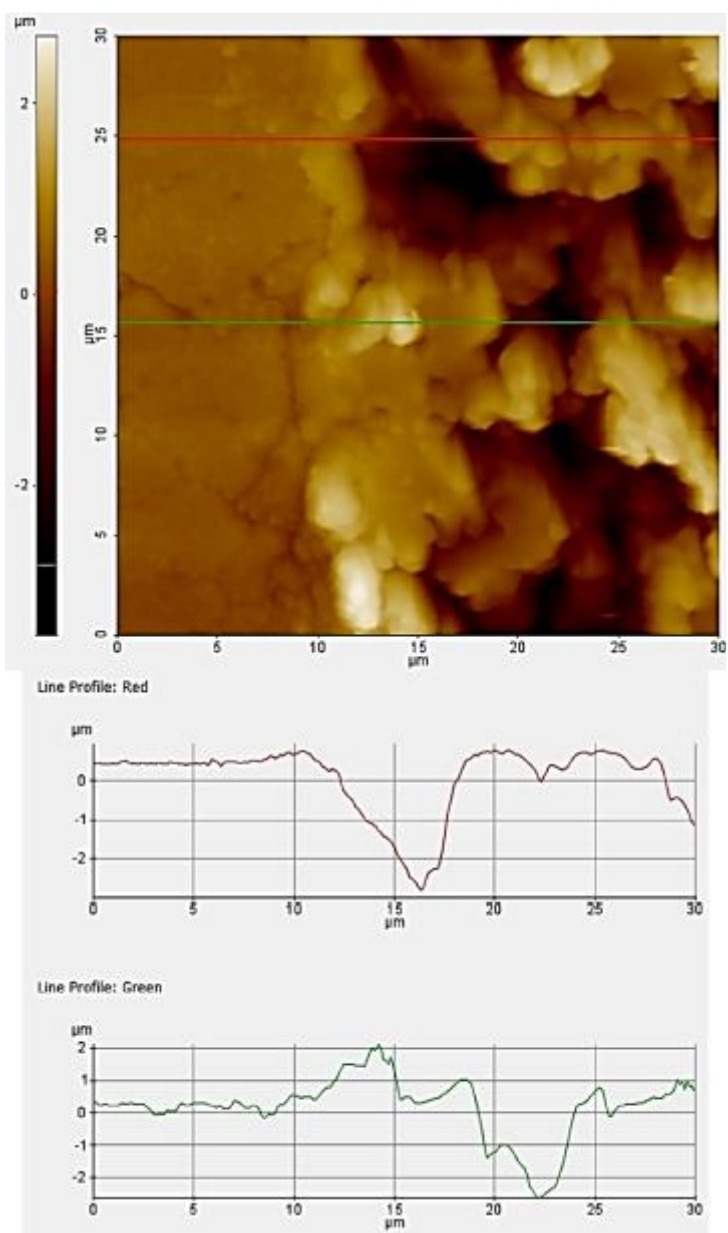

Fig. 5. AFM profile on the border between ablated and non-ablated region of the NPs-PDMS nanocomposites.

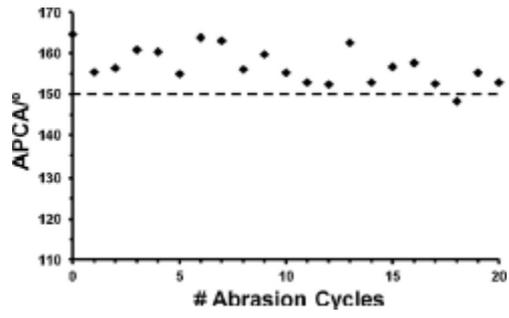

Fig. 6. APCA of water plotted against the number of abrasion cycles. The abrasion length was 4 cm.

Table 1. Analysis of the % elemental composition by XPS data of bare PDMS, laser-treated PDMS, nanocomposite without laser treatment, laser-treated nanocomposite, iron nanopowder. The laser-treated samples were irradiated with 30 pulses of 0.8 J/cm2 laser fluence.

| Material element | PDMS | PDMS laser-treated | Nanocomposite | Nanocomposite laster-treated | NPs |
|---|---|---|---|---|---|
| O1s | 19.42 | 20.99 | 18.79 | 25.05 | 6.37 |
| C1s | 46.85 | 43.07 | 46.77 | 37.14 | 92.99 |
| Si2p | 33.72 | 35.94 | 34.44 | 37.81 | - |
| Fe2p | - | - | - | - | 0.65 |

The ablated films were also found to be durable after 20 cycles of linear mechanical abrasion. Fig. 6 shows the APCA plotted against the number of abrasion cycles. In particular, a cylinder of 2.5 cm diameter with approximate weight 100 g was moved linearly on the surface with a velocity of 5 mm/s. The abrasion length was 4 cm and the APCA was measured after the end of each abrasion cycle. The films retained their high values of APCA even after 20 abrasion cycles. The elasticity of the PDMS matrix seemed to prevent the formation of scratches on the film's surface that was found to be wear resistant. This kind of wear resistant behavior reported here is pretty much consistent with the observations that Yuan et al. made on superhydrophobic honeycomb-like PDMS structures [50].

Preliminary experiments have shown that the laser treated elastomeric films can be utilized for water–oil separation applications. Specifically the superhydrophobic/superoleophilic magnetic films are rendered porous through the fabrication of laser-drilled apertures of different diameters. Fig. 7 shows an SEM image of a single pore with 120 μm diameter while other pores of diameters 160 μm, 200 μm and 300 μm are formed in the same way. The laser-drilled films were left to float on water and ethanol mixtures to test their filtering capability (Fig. 7). When the sample was floating on water, no penetration could be observed through the pores of all sizes. However, in contact with ethanol, liquid penetrating from the side of the sample that was in contact with the liquid was observed in all cases. As shown in Fig. 7, after 30 s there is a clear difference in the quantity of ethanol that passes from one side of the sample to the other due to the varying size of the pores. The pores with bigger diameter allowed more liquid to pass to the other side.

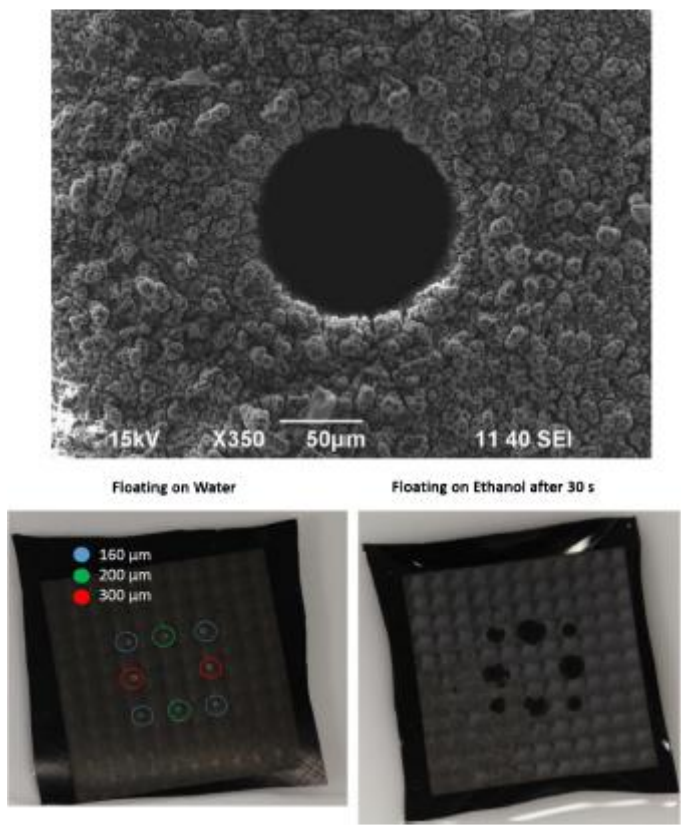

Fig. 7. Top: SEM image of a laser-drilled aperture of 120 μm diameter. Bottom: Demonstration of the porosity size effect for water and ethanol as the samples are floating on these two liquids. While floating on water there is no penetration of liquid even through the biggest pores (300 μm diameter). While floating on ethanol after 30 s can be clearly observed some liquid penetrating through the pores, with the bigger pores transmitting more liquid.

## 4. Conclusions

We report the development and characterization of magnetic superhydrophobic and superoleophilic nanocomposite thin sheets by laser ablation. The laser ablation induced chemical and structural changes (both to micro- and nanoscale) to the surface of the nanocomposites that were rendered superhydrophobic and superoleophilic. The use of NPs was found to facilitate the ablation process, since the number of pulses and the laser fluence required for the ablation were greatly reduced compared to the ablation of the bare polymer, and moreover, to provide to PDMS magnetic properties. Such nanocomposite films can be incorporated in magnetic MEMS for the controlled separation of liquids or used like switches. Moreover due to their superhydrophobic/superoleophilic properties they can be considered as a potential candidate material for oil–water separation applications.

**Appendix A. Supplementary data**

Supplementary data associated with this article can be found, in the online version.